\documentclass[runningheads]{llncs}
\usepackage[T1]{fontenc}
\usepackage{graphicx}
\usepackage{bm}
\usepackage{amsmath}
\usepackage{amssymb}
\usepackage{amsfonts}
\usepackage{dsfont}
\usepackage{algorithmic}
\usepackage{algorithm2e}
\usepackage{cleveref}

\begin{document}
\title{Optimising Inpainting Data with\\Delaunay Averages}
%
%\titlerunning{Abbreviated paper title}
% If the paper title is too long for the running head, you can set
% an abbreviated paper title here
%
\author{Vassillen Chizhov
\and
Joachim Weickert}%\inst{1} \orcidID{0000-0001-7082-6367} \orcidID{0000-0002-8494-0045}
\authorrunning{V.\ Chizhov and J.\ Weickert}
% First names are abbreviated in the running head.
% If there are more than two authors, 'et al.' is used.
%
\institute{Mathematical Image Analysis Group, Faculty of 
Mathematics and Computer Science, Saarland University, 
66041 Saarbr\"ucken, Germany
\email{\{chizhov,weickert\}@mia.uni-saarland.de}\\
\url{https://www.mia.uni-saarland.de/index.shtml}}
\maketitle              % typeset the header of the contribution
%
%-----------------------------------------------------------------------------
\begin{abstract}
%The abstract should briefly summarize the contents of the paper in
%150--250 words.
%
Inpainting-based image compression usually stores an optimised subset 
of all pixel locations and their colour values. In the decoding
phase, the missing data are approximated via inpainting. Since the
reconstruction quality depends critically on the selection of the 
stored data, we introduce a novel feature type: We store the vertex 
locations of a Delaunay triangulation together with the average colour 
values inside all triangles. We show that combining this feature type 
with homogeneous diffusion inpainting creates an elegant mathematical 
formulation with a positive definite linear system of equations. 
Even a simple solver such as the conjugate gradient method allows
the handling of large images.  
To make our Delaunay averages maximally adaptive to the image, we 
develop an efficient data optimisation strategy specifically tailored 
to them. It incorporates ideas successfully used in the stippling 
literature. Experiments show that our approach outperforms the 
popular inpainting with optimised colour values by a large margin.
Last but not least, we discover a favourable scaling behaviour: 
Doubling the image resolution allows us to halve the percentage of 
stored data while maintaining the quality level. This is attractive 
for compressing modern high-resolution images, where even data 
densities below 1 \% yield appealing reconstructions.

\keywords{Inpainting  \and Image Compression \and Partial Differential
Equations (PDEs) \and Delaunay triangulation.}
\end{abstract}

%%%%%%%%%%%%%%%%%%%%%%%%%%%%%%%%%%%%%%%%%%%%%%%%%%%%%%%%%%%%%%%%%%%%%%%%%%%%

\section{Introduction}

Inpainting refers to the reconstruction of an image from a subset 
of its data~\cite{GL14}. It has also emerged as a powerful 
paradigm for lossy image compression (see e.g.~\cite{GWWB08,SPME14}),
offering an interesting and conceptually simple alternative to 
trans\-form-based standards such as JPEG and its successors.
Inpainting-based codecs consist of an encoding and a decoding stage. 
During encoding, they select and store an optimised subset of the image 
data -- typically a small fraction of pixel locations (often referred 
to as the inpainting \textit{mask}) together with their greyscale 
or colour values. In the decoding phase, the missing image content is 
reconstructed from these sparse data via an inpainting process.
Even simple inpainting methods such as homogeneous diffusion can achieve 
remarkably high reconstruction quality -- provided that the stored data 
are carefully optimised~\cite{CRP14,MHWT12,SPME14}. 

As we will see below, there has been extensive research on optimising 
the locations of the selected pixels. The potential of alternative 
feature types, however, has not been fully explored so far: Using edge 
information is well-suited for cartoon-like or piecewise smooth images 
(see e.g.~\cite{Ca88,MBWF11}), but loses it efficiency for more general 
imagery. Several authors have also reported some success with 
derivative features~\cite{BBG15,SPHW16}.

More recently, Jost et al.~\cite{JCW23} have established a general
framework that allows to incorporate multiple features which can be
expressed by linear constraint equations. This includes e.g.~pointwise 
greyscale or colour values, derivatives, but also novel integral 
features that compute local averages over a {\em fixed} neighbourhood. 
Their performance benefits from the robustness of integration. A unifying 
characteristic of all features in \cite{JCW23} is their shift invariance: 
They can be expressed as convolutions and, thus, do not adapt themselves 
to the spatial configuration of the inpainting mask. For this reason, 
we refer to them as \textit{non-adaptive} features. Adaptive features 
that exploit the mask configuration have not been studied yet.

%----------------------------------------------------------------------------

\medskip
\noindent {\bf Our Contributions.}
While our paper builds upon the success of integral features, we 
improve them by introducing the first \textit{adaptive} feature.
This offers a number of advantages. Our main contributions are as 
follows:

\vspace{-1mm}
\begin{enumerate}
\item The proposed feature specifies the average colour value for each 
      triangle of a Delaunay partitioning. These colour values are stored 
      together with the vertex locations of the Delaunay triangulation. 
      Since Delaunay triangulations are fully determined by their vertices
      and efficient algorithms exist that construct them from these 
      points~\cite{AKL13}, no connectivity information is stored.

\medskip
\item We show that equipping homogeneous diffusion inpainting with this 
      Delaunay feature leads to an elegant discrete formulation. The 
      resulting linear system of equations is positive definite. 
      Already a simple conjugate gradient method~\cite{GVC13} can 
      handle large image resolutions in practice.

\medskip
\item By optimising the positions of the Delaunay vertices, we obtain 
      a nonlinear feature adaptation to the mask. We develop a data 
      optimisation strategy that is specifically tailored to our 
      feature type. It incorporates ideas from the Linde-Buzo-Gray 
      algorithm as is used e.g.~in the stippling literature~\cite{DSZ17}.

\medskip
\item Experiments on natural images illustrate that Delaunay features 
      can outperform the popular pointwise colour features by a large
      margin.

\medskip
\item We discover a favourable scaling behaviour that makes our approach 
      particularly attractive for compressing high-resolution images. 
      Here high quality reconstructions are possible even with mask 
      densities below $1 \%$.
\end{enumerate}

\vspace{-1mm}
\noindent
In our paper we do not discuss questions of efficient encoding of the 
selected data with additional strategies such as entropy coding and data 
quantisation. These problems are nontrivial and are analysed in detail 
elsewhere~\cite{MPW21}.

%-----------------------------------------------------------------------------

\medskip
\noindent\textbf{Related Work.}
% Our goal in this paper is to replace the pointwise colour values used in 
% classical inpainting-based compression by a more expressive feature type. 
As already mentioned, alternative features considered so far 
include edges~\cite{Ca88,MBWF11}, derivatives~\cite{BBG15,SPHW16}, and 
local averages over a fixed neighbourhood~\cite{JCW23}. These are all 
non-adaptive features, whereas the proposed Delaunay feature is adaptive.

Outside the image compression community, ways have been explored to 
reconstruct an image from its zero-crossings~\cite{ZR86}, 
top points in scale-space~\cite{KLDJ05}, junctions~\cite{CCM97}, and 
SIFT descriptors~\cite{WJP11}. While these approaches are interesting, 
their performance is not competitive for inpainting-based compression.

The work of Jost et al.~\cite{JCW23} is particularly relevant for us, 
since its framework covers general features that can be expressed as 
linear equality constraints. We propose a simpler way to incorporate 
our Delaunay constraint. While the formulation in \cite{JCW23} 
creates indefinite systems of equations, we end up with positive definite 
ones. They are numerically better suited for handling larger images.

High reconstruction quality requires a careful placement of the
stored features \textit{(spatial optimisation)}. Our strategy follows 
and refines ideas from Jost et al.~\cite{JCW23}. As their spatial 
optimisation algorithm is restricted to non-adaptive features, 
we have to extend it in a non-trivial way. In particular, since our 
approach relies on averages over a partition, we adopt a region 
splitting strategy inspired by the Linde–Buzo–Gray algorithm 
used e.g.~for stippling~\cite{DSZ17}.

Beyond this line of work, a variety of alternative approaches to 
spatial optimisation have been explored in the literature. These 
include analytic methods~\cite{BBBW09}, non-smooth optimisation 
techniques~\cite{BLPP17,HSW13,OCBP14}, neural networks~\cite{PSAW23,SPKW23}, 
probabilistic sparsification~\cite{MHWT12}, 
and densification strategies~\cite{DAW21,KBPW18}. While our work 
incorporates densification concepts, the remaining approaches are 
not directly applicable to our setting without substantial adjustments.

Delaunay triangulations have repeatedly proven to be a key component
for high-quality inpainting-based compression methods~\cite{CW21,DDI06},
where they perform linear spline interpolation of the mask data. These
applications, however, create nondifferentiable results. This is in
contrast to our setting where they provide integral constraints for
a diffusion equation with strong regularising properties.

%-----------------------------------------------------------------------------

\medskip
\noindent\textbf{Paper Structure.}
In \Cref{sec:homdiff}, we review homogeneous diffusion inpainting. 
The Delaunay feature type is introduced and embedded in the inpainting
setting in \Cref{sec:delaunay}. \Cref{sec:spatial_optimisation} describes 
our spatial optimisation strategy. In \Cref{sec:experiments} we present 
experimental evaluations, and we conclude our paper in \Cref{sec:conclusions}.

%%%%%%%%%%%%%%%%%%%%%%%%%%%%%%%%%%%%%%%%%%%%%%%%%%%%%%%%%%%%%%%%%%%%%%%%%%%%%

\section{Review of Homogeneous Diffusion Inpainting}
\label{sec:homdiff}

In our paper we use homogeneous diffusion inpainting~\cite{Ca88}, which 
is one of the most popular strategies in inpainting-based compression. 
This linear operator offers many advantages: It is simple, 
parameter-free, permits highly efficient algorithms, and it is 
mathematically better understood than other inpainting methods. 
Since it yields a surprisingly good performance when being equipped 
with highly optimised 
data~\cite{BBBW09,BLPP17,Ca88,CW21,HSW13,JCW23,KCW25,MBWF11,MHWT12,OCBP14,PSAW23,SPKW23}, 
it has become a standard for data optimisation in inpainting.
We briefly review its main concepts.

%............................................................................
%
\begin{figure}
\setlength{\tabcolsep}{2mm}
\centering
    \begin{tabular}{ccc}
    \includegraphics[width=0.3\textwidth]{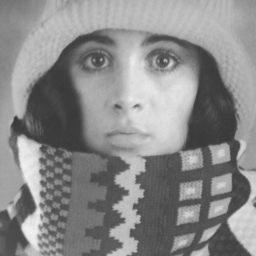}
    &
    \includegraphics[width=0.3\textwidth]{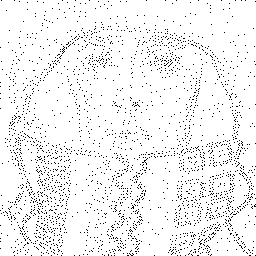}
    &
    \includegraphics[width=0.3\textwidth]{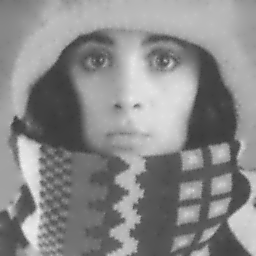}  
    \\
    [1mm]
   \small{original image $\bm{f}$} 
   &
   \small{mask visualised as $\bm{1}\!-\!\bm{c}$} 
   &
   \small{inpainting $\bm{u}$} 
    \end{tabular}
    \caption{Homogeneous diffusion inpainting with a 
    mask with $5\,\%$ data density.}
    \label{fig:trui_harmonic}
\end{figure}
%
%---------------------------------------------------------------------------

\medskip
\noindent {\bf Continuous Formulation.}
Consider a continuous greyscale image $f : \Omega \rightarrow \mathbb{R}$ 
defined on a rectangular domain $\Omega$. For RGB colour images, one
may proceed channel by channel. Instead of storing the full image over 
$\Omega$, we restrict the information to a subset $K \subset \Omega$, 
referred to as the \emph{data domain} (or mask pixels in the discrete setting). 
The unknown image values on $\Omega \setminus K$ are reconstructed by 
solving the Laplace equation with reflecting boundary conditions on 
$\partial \Omega$:
\begin{alignat}{3}
 -\Delta u(\bm{x}) &\,=\, 0, \quad &\bm{x} &\in \Omega \setminus K, \\ 
 u(\bm{x}) &\,=\, f(\bm{x}), \quad &\bm{x} &\in K, \label{eq:ic}\\ 
 \partial_{\bm{n}} u(\bm{x}) &\,=\, 0, \quad &\bm{x} &\in \partial \Omega,
\end{alignat}
where \( \bm{n} \) denotes the outward normal to the boundary, and 
$\Delta = \partial_{xx} +\partial_{yy}$ is the Laplacian in 2D.
For a more compact representation, we introduce a confidence function
$c(\bm{x})$ that specifies the locations of the stored data:
\begin{equation}
    c(\bm{x}) \,=\, 
%    \mathds{1}_K(\bm{x}) =
    \begin{cases}
        1, & \bm{x} \in K, \\
        0, & \bm{x} \notin K.
    \end{cases}
\end{equation}
It allows to rewrite the continuous inpainting problem as 
\begin{alignat}{3}
 \label{eq:harmonic_inp_cont_compact}
 \bigl(c(\bm{x}) + (1 - c(\bm{x}))(-\Delta)\bigr)\,u(\bm{x}) 
   &\,=\, c(\bm{x}) f(\bm{x}), &\quad& \bm{x} \in \Omega,\\
 \label{eq:harmonic_inp_cont_compact2}
\partial_{\bm{n}} u(\bm{x}) &\,=\, 0, &\quad& \bm{x} \in \partial \Omega.
\end{alignat}
%
%---------------------------------------------------------------------------
%
{\bf Discrete Formulation.}
Discretising \( \Omega \) on a regular grid yields vector-valued 
counterparts \( \bm{u},\, \bm{f} \in \mathbb{R}^N \) and the inpainting 
\textit{mask} \( \bm{c} \in \{0,1\}^N \), where the value $1$ 
indicates a known data point, and $0$ a location to be inpainted. 
The Laplace operator with reflecting boundary conditions is 
approximated by the standard five-point stencil~\cite{MM05}, 
resulting in a positive semi-definite matrix \( \bm{L} \approx -\Delta \). 
With the diagonal matrix \( \bm{C} = \operatorname{diag}(\bm{c}) \), the 
discrete counterpart to 
(\ref{eq:harmonic_inp_cont_compact})--(\ref{eq:harmonic_inp_cont_compact2}) 
becomes
\begin{equation}
    \label{eq:discrete_harmonic_inpainting}
    (\bm{C} + (\bm{I}\!-\!\bm{C})\,\bm{L})\,\bm{u} \,=\, \bm{C} \bm{f}.
\end{equation}
This linear system admits a unique solution $\bm{u}$ whenever the mask $\bm{c}$
is non-empty~\cite{MBWF11}. \Cref{fig:trui_harmonic} illustrates the 
reconstruction for a classical test image. A closer inspection 
of the inpainted image reveals the presence of logarithmic singularities, 
which are intrinsic to homogeneous diffusion when pointwise data are 
specified -- these singularities are inherited from the continuous 
setting. 
They can be avoided, for instance, by imposing integral features over 
some area instead of pointwise ones~\cite{JCW23}. Also our adaptive 
Delaunay feature is an integral feature.

%%%%%%%%%%%%%%%%%%%%%%%%%%%%%%%%%%%%%%%%%%%%%%%%%%%%%%%%%%%%%%%%%%%%%%%%%%%%%

\section{Our Adaptive Feature Type with Delaunay Averages}
\label{sec:delaunay}

In this section, we construct an adaptive integral feature that serves as 
a replacement for the classical pointwise interpolation constraints 
$\bm{C}\bm{u} = \bm{C}\bm{f}$ from \eqref{eq:ic} and 
\eqref{eq:discrete_harmonic_inpainting}. 
It is based on a Delaunay triangulation.

%-----------------------------------------------------------------------------

\medskip
\noindent {\bf Averages over a Partition.}
We consider a partition 
$\mathcal{P} = \{\mathcal{R}_1,\ldots,\mathcal{R}_M\}$
of the image domain $\Omega$, and prescribe the averages of $u$ over 
each region in $\mathcal{P}$. To this end, we introduce a matrix 
$\,\bm{P}=(p_{i,j}) \in \mathbb{R}^{N \times N}\,$ with 
\begin{equation}
  \label{eq:projector_closed_form}
  p_{i,j} \,=\,
    \begin{cases}
        \frac{1}{|\mathcal{R}_k|}, & \text{if pixels $i$ and $j$ belong to 
        the same region } \mathcal{R}_k, \\
        0, & \text{otherwise}.
    \end{cases}
\end{equation}
The matrix--vector product $\widetilde{\bm{f}} = \bm{P} \bm{f}$ 
yields an image that is piecewise constant on $\mathcal{P}$. In each 
region $\mathcal{R}_k$, the result $\widetilde{\bm{f}}$ takes the average 
value of $\bm{f}$ over that region. 
$\bm{P}$ is an orthogonal projection matrix (i.e.~$\bm{P}^2\!=\!\bm{P}$ and 
$\bm{P}^\top\!=\!\bm{P}$) onto the subspace of 
functions that are constant on each region of $\mathcal{P}$.
% Equivalently, $\widetilde{\bm{f}}$ is the best $L_2$ approximation of 
% $\bm{f}$ onto the subspace of functions that are constant on each region 
% of $\mathcal{P}$. 

%-----------------------------------------------------------------------------

\medskip
\noindent {\bf Projection-Based Inpainting Formulation.}
To derive a reformulation of the inpainting problem that enforces
average constraints over the partition $\mathcal{P}$, we revisit
\eqref{eq:discrete_harmonic_inpainting}. We observe that the matrix
$\bm{C}$ acts as an orthogonal projector in the classical setting,
while the discrete Laplace equation $\bm{L}\bm{u} = \bm{0}$ 
is solved on its orthogonal complement $\,\bm{I}\!-\!\bm{C}$.
Motivated by this observation, we propose the following generalised
projection-based inpainting formulation:
\begin{equation}
  \label{eq:formulation_projector}
  (\bm{P} + (\bm{I}\!-\!\bm{P})\, \bm{L})\,\bm{u} \,=\, \bm{P} \bm{f}.
\end{equation}
This extension of \eqref{eq:discrete_harmonic_inpainting}, 
with $\bm{C}$ being replaced by $\bm{P}$, offers several advantages:
\begin{itemize}
\item This substitution is straightforward only because a closed-form 
      expression for $\bm{P}$ is available; see 
      \eqref{eq:projector_closed_form}. Without such a representation, 
      one would typically resort to the saddle-point
      formulation in~\cite{JCW23}. It leads to an indefinite system 
      which requires relatively expensive numerical solvers that do
      not scale well.
      In our setting, $\bm{L}$ is a discretisation of the negative
      Laplacian, which ensures that the system matrix in
      \eqref{eq:formulation_projector} is symmetric positive definite.
      Consequently, classical solvers such as the conjugate
      gradient method are guaranteed to converge~\cite{GVC13}. 
      Its scaling behaviour is more favourable and allows to handle 
      also large image resolutions in practice.

\medskip
\item Our reconstruction $\bm{u}$ from \eqref{eq:formulation_projector} 
      is constrained to reproduce  the average grey value in each
      Delaunay triangle of the partition. Therefore, it has the same 
      average grey value as the original image $\bm{f}$. This is an 
      advantage over inpainting with pointwise data constraints which 
      lacks this property.

\medskip
\item The fact that \eqref{eq:formulation_projector} aims at 
      good reconstructions over image subareas rather than exact ones 
      at individual mask points makes it similar in spirit to tonal 
      optimisation approaches~\cite{MHWT12}. They optimise the grey 
      values at mask pixels in a postprocessing step to minimise the 
      global reconstruction error. While this is computationally fairly 
      costly, \eqref{eq:formulation_projector} may serve as an
      inexpensive alternative. 
\end{itemize}

%-----------------------------------------------------------------------------

\medskip
\noindent {\bf Mask Adaptivity.}
To adjust the partition $\mathcal{P}$ on the mask $\bm{c}$ and 
obtain an adaptive feature, we use a geometric construction 
derived solely from the sparse set of mask pixels. 
This avoids the need to store additional connectivity information.

\begin{figure}[tb]
\setlength{\tabcolsep}{2mm}
\begin{center}
    \begin{tabular}{ccc}
    \includegraphics[width=0.3\textwidth]{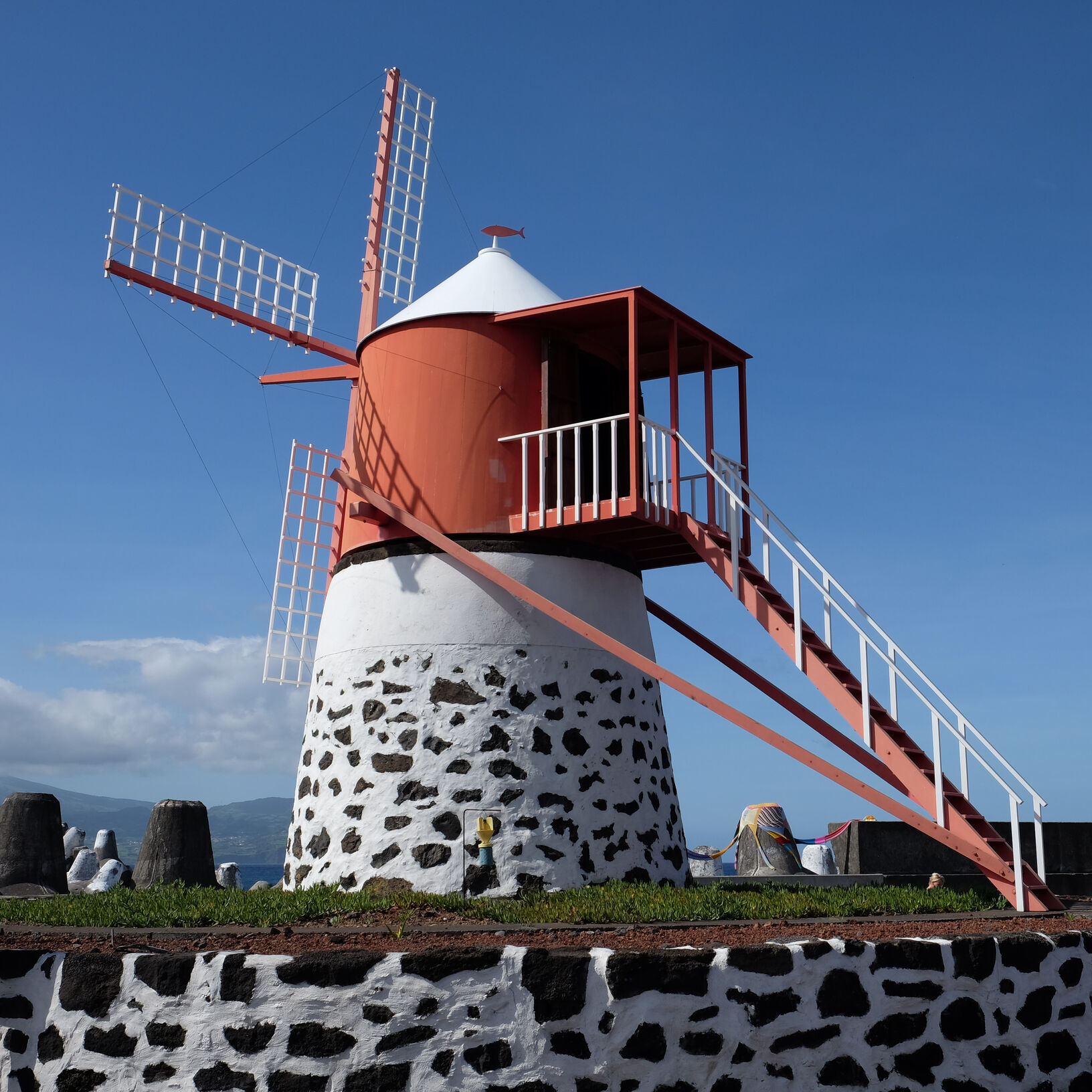}
    &
    \includegraphics[width=0.3\textwidth]{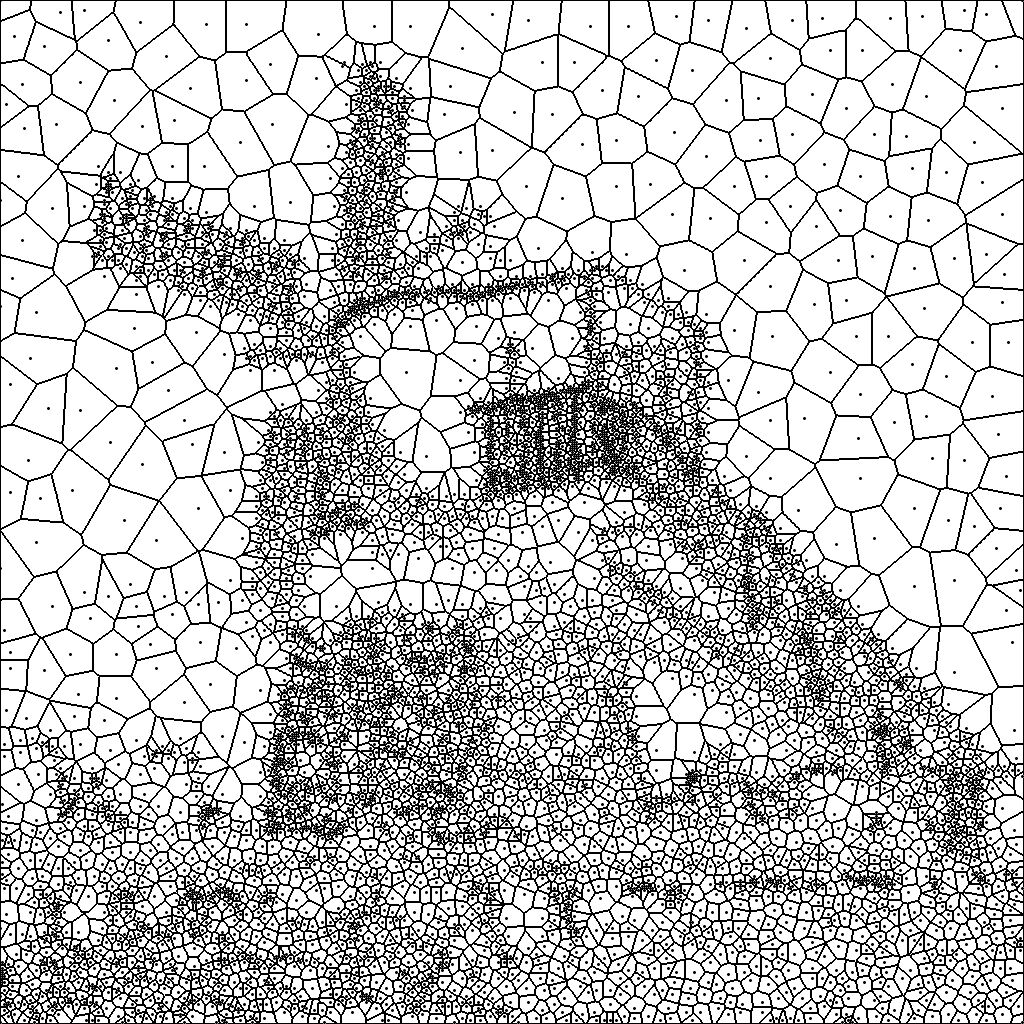}
    &
    \includegraphics[width=0.3\textwidth]{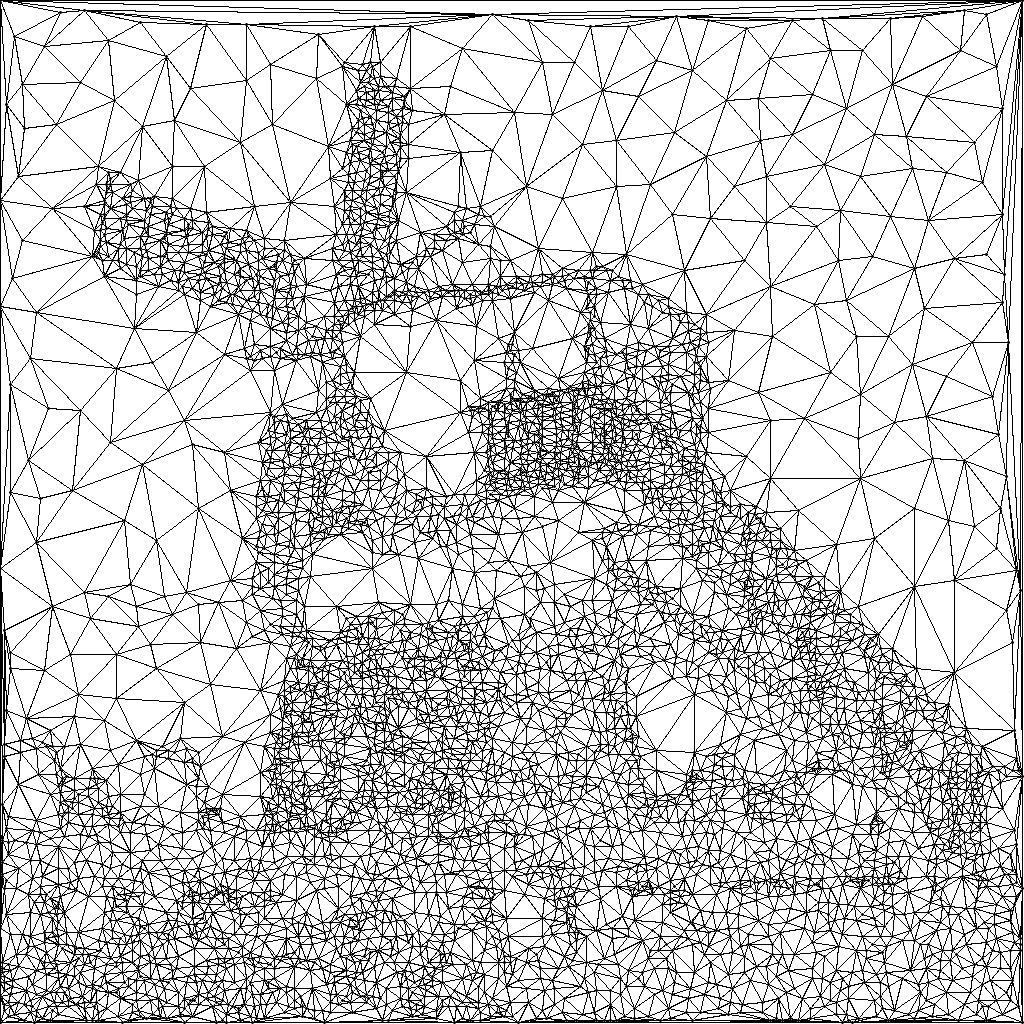}  
    \\
    [1mm]
   \small{\emph{windmill}}
   &
   \small{Voronoi diagram}
   &
   \small{Delaunay triangulation}
    \end{tabular}
\end{center}
    \caption{Voronoi diagram and its dual Delaunay triangulation for 
    a mask used to reconstruct the \textit{windmill}\, image. Photo by 
    J.~Weickert.}
    \label{fig:Delaunay}
\end{figure}

Two common structures of this type are the Voronoi diagram~\cite{AKL13} 
and the Delaunay triangulation~\cite{AKL13}. A {\em Voronoi diagram} 
partitions the domain into regions consisting of all points closest to 
a given mask point. The {\em Delaunay triangulation} is its dual graph: 
Mask points corresponding to adjacent Voronoi regions are connected by edges, 
forming a triangulation; see \Cref{fig:Delaunay} for an illustration.

Based on its superior empirical performance, we use the Delaunay 
triangulation. The Voronoi diagram appears in the data optimisation 
stage. Our Delaunay-based feature is adaptive, in the sense that 
both the shape and density of the triangles over which averages are 
prescribed depend on the mask and thus on the image; see \Cref{fig:Delaunay}. 
This stands in contrast to the non-adaptive features considered 
in~\cite{JCW23}, such as averages over disks of fixed radius. 

The adaptivity of the feature makes it a nonlinear function of the 
mask $\bm{c}$. It requires a specialised spatial optimisation strategy 
that we discuss next.  

%..........................................................................

%%%%%%%%%%%%%%%%%%%%%%%%%%%%%%%%%%%%%%%%%%%%%%%%%%%%%%%%%%%%%%%%%%%%%%%%%%%%%

\section{Spatial Optimisation}
\label{sec:spatial_optimisation}

Our proposed feature requires a dedicated spatial optimisation strategy. 
Among existing methods for classical inpainting, densification approaches 
based on error maps~\cite{KBPW18} over Voronoi diagrams~\cite{DAW21} 
offer a favourable balance between reconstruction quality and 
computational cost. We therefore retain these key ideas and adapt them 
to our setting. We consider a region-splitting strategy, rather than 
introducing new mask points into an otherwise fixed configuration, as 
is common in previous approaches. The regions to be split are the 
Voronoi cells around the mask pixels -- the dual of our Delaunay 
triangulation. 
The resulting procedure is summarised in \Cref{alg:voronoi-densification}. 
The initial mask is constructed from points of a low-discrepancy sequence 
based on the golden ratio~\cite{Ro18}. 
During the densification process, Voronoi regions are split along the 
principal eigenvectors of the covariance matrices of the error within 
their associated regions, with displacements proportional to the areas 
of those regions.
These design choices are well motivated and draw inspiration from the 
Linde--Buzo--Gray algorithm as used in the stippling literature~\cite{DSZ17}.

%-----------------------------------------------------------------------------

\begin{algorithm}[tb]
\textbf{Input:} Original image $\bm{f}$,  
    number of iterations $n$, number of desired mask points $m$  \\
\textbf{Output:} Inpainting mask $\bm{c}^n$, reconstruction $\bm{u}$\\
\textbf{Initialise:} Initial mask $\bm{c}^1$ with 
    $\lceil \frac{m}{n} \rceil$ mask pixels\\
% \textbf{Loop:}\\
    \begin{algorithmic}[1]
        \FOR{$k=1$ \TO $n-1$}
        % \FOR{$i=1, \dots ,n-1$}
            \STATE
            Construct the Voronoi diagram $\{\mathcal{V}_{j}\}$ 
            of the current mask pixels.
            \STATE
            Compute the inpainting 
            $\bm{u}^k = \bm{u}(\bm{c}^k,\bm{f})$
            and the error map $\bm{e}^k=\bm{u}^k-\bm{f}$. 
            \STATE
            Compute the cells squared $2$-norm errors:
            $\forall j, \, e^k_{\mathcal{V}_j} = \sum_{i\in \mathcal{V}_j} (e^k_i)^2$.
            \STATE
            Find the $\lceil\frac{m}{n}\rceil$ Voronoi cells
            $\{\mathcal{V}_{j_i}\}_{i=1}^{\lceil \frac{m}{n} \rceil}$ with the 
            largest 
            errors $\{e^k_{\mathcal{V}_{j_i}}\}_{i=1}^{\lceil\frac{m}{n}\rceil}$.
            \STATE
            For each cell in $\{\mathcal{V}_{j_i}\}_{i=1}^{\lceil\frac{m}{n}\rceil}$ 
            split it into two cells   
            along the principal axis of the covariance matrix of 
            the error map restricted to the cell. The offset is chosen  
            proportional to the area of the cell.
        \ENDFOR
 \end{algorithmic}
    \vspace{4mm}
 	\caption{Voronoi densification for the Delaunay feature.
        (The number of split Voronoi cells can be adjusted slightly to 
        reach exactly $m$ mask points.)
        } 
	\label{alg:voronoi-densification} 
\end{algorithm}

%%%%%%%%%%%%%%%%%%%%%%%%%%%%%%%%%%%%%%%%%%%%%%%%%%%%%%%%%%%%%%%%%%%%%%%%%%%%%

\section{Experiments}
\label{sec:experiments}

\begin{figure}[!tbh]
\centering
\setlength{\tabcolsep}{2mm}
\begin{tabular}{ccc}
\small{\textit{boats}} &
\small{\textit{elpaso}} &
\small{\textit{flowers}} \\[1.0mm]
\includegraphics[width=0.3\textwidth]{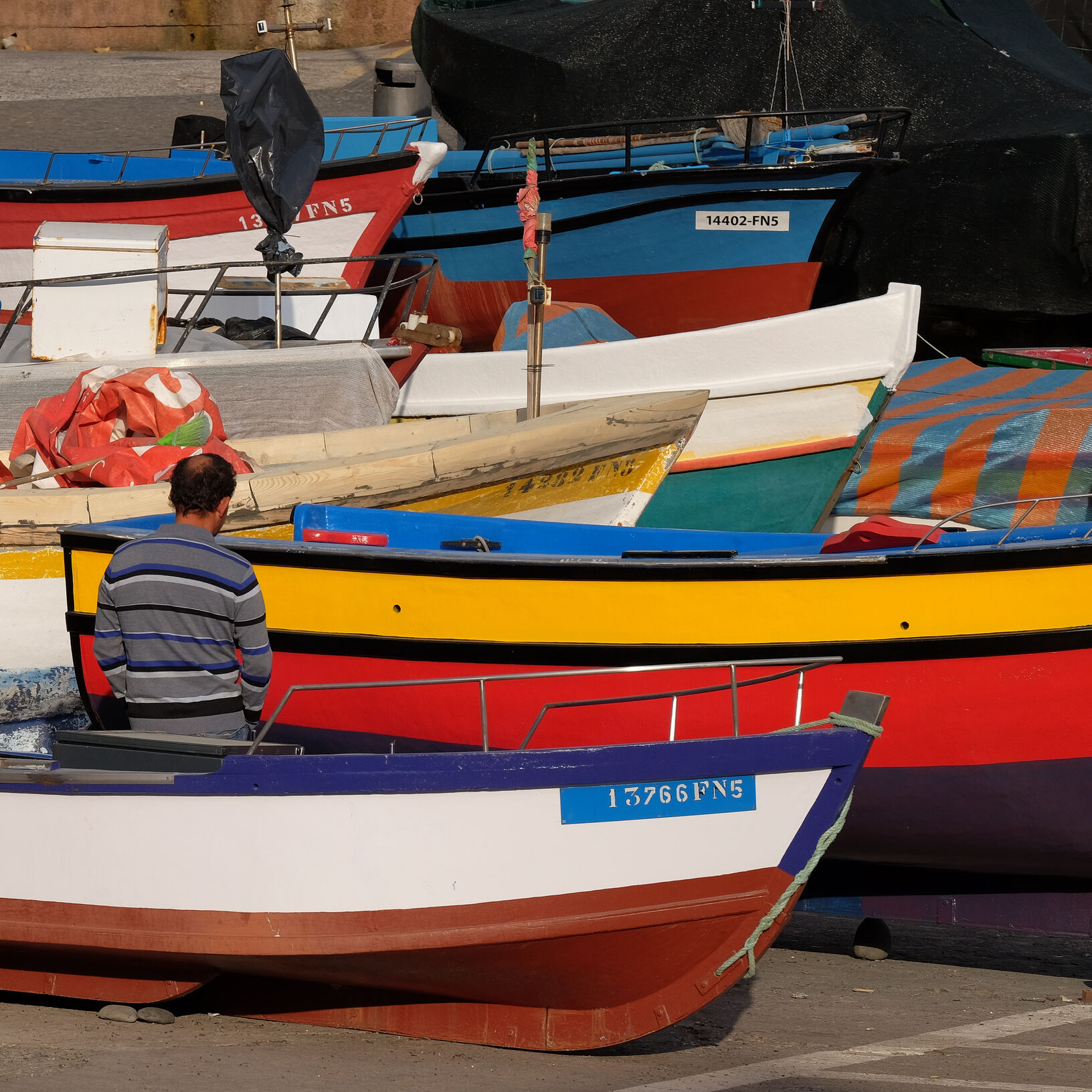} &
\includegraphics[width=0.3\textwidth]{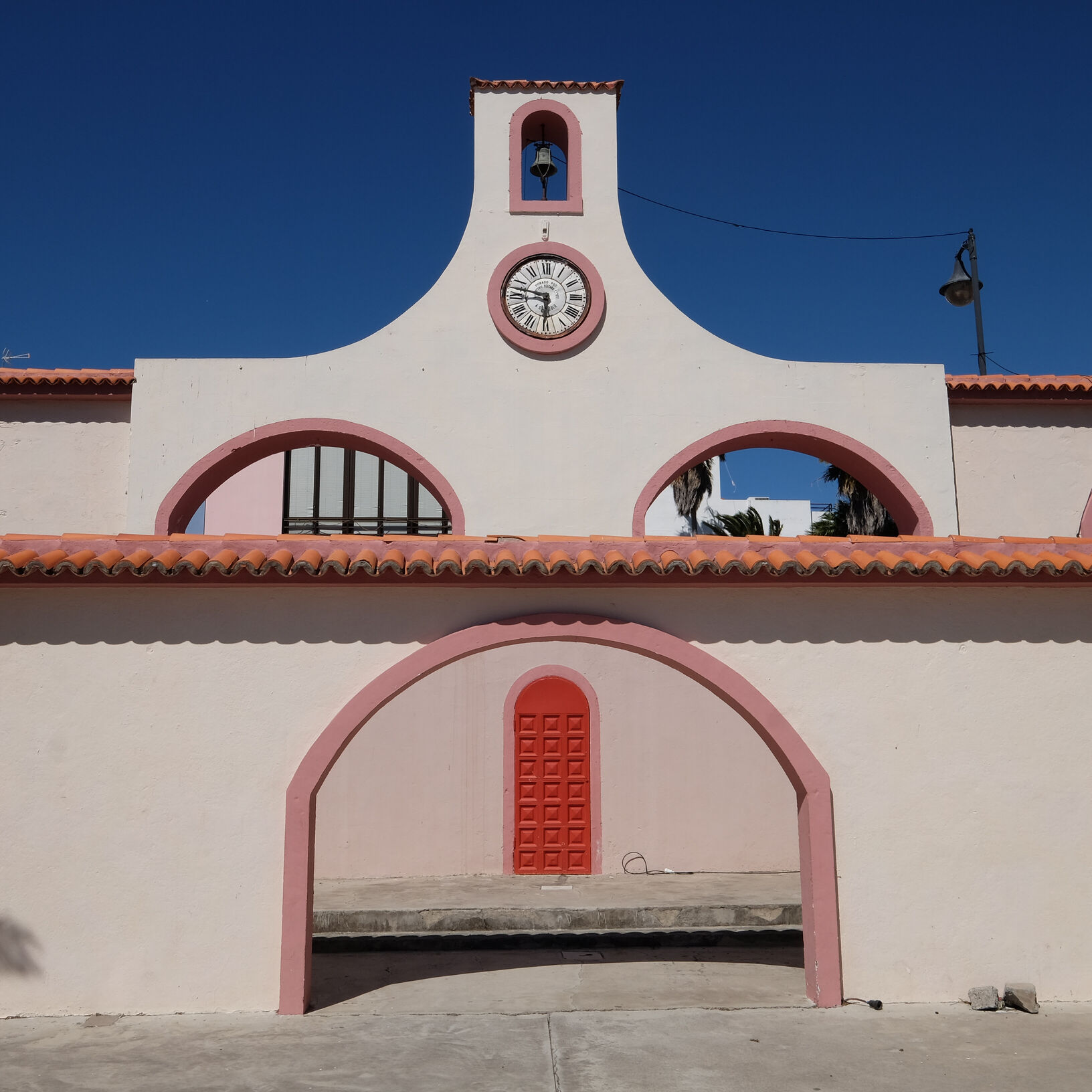} &
\includegraphics[width=0.3\textwidth]{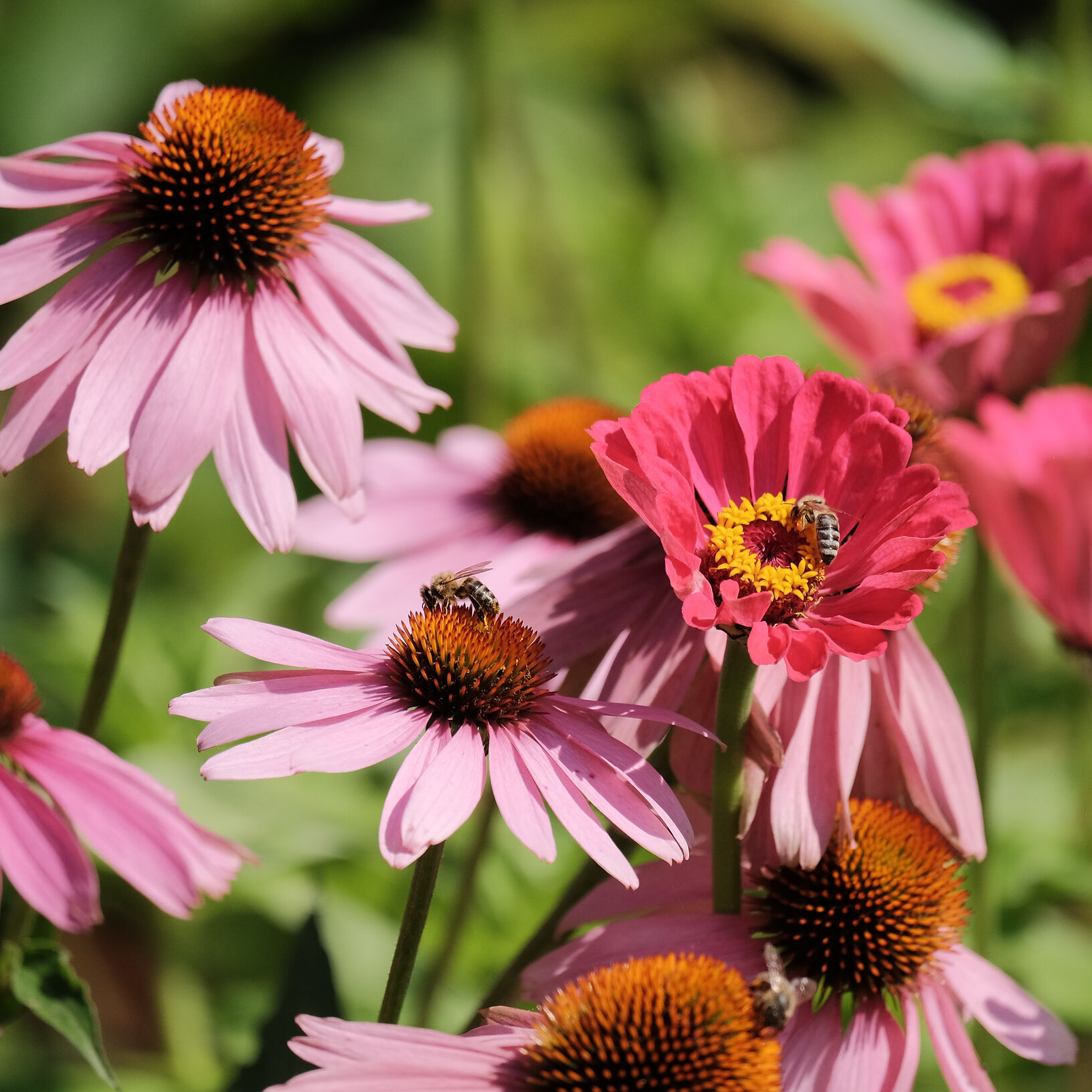}\\[3mm]
\includegraphics[width=0.3\textwidth]{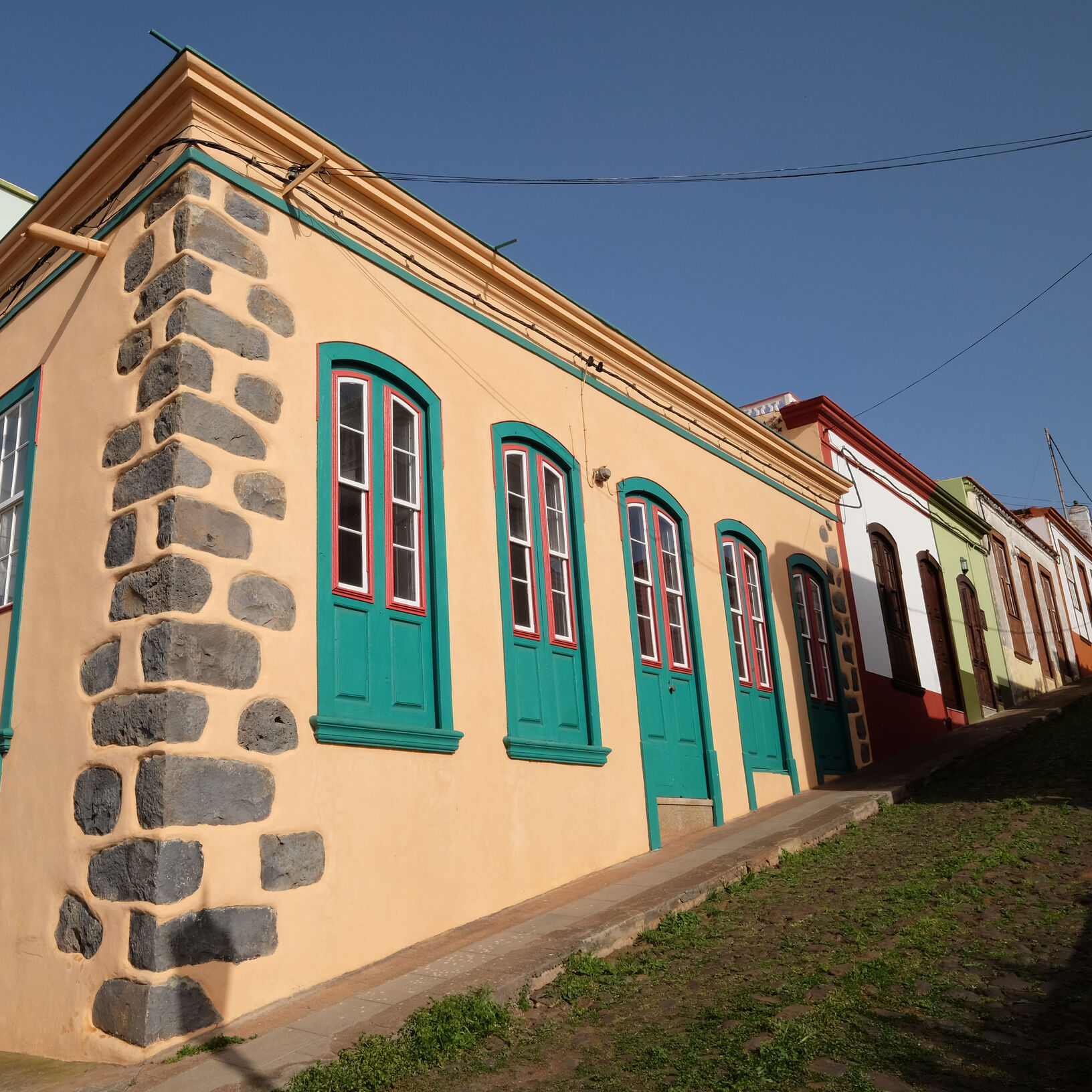} &
\includegraphics[width=0.3\textwidth]{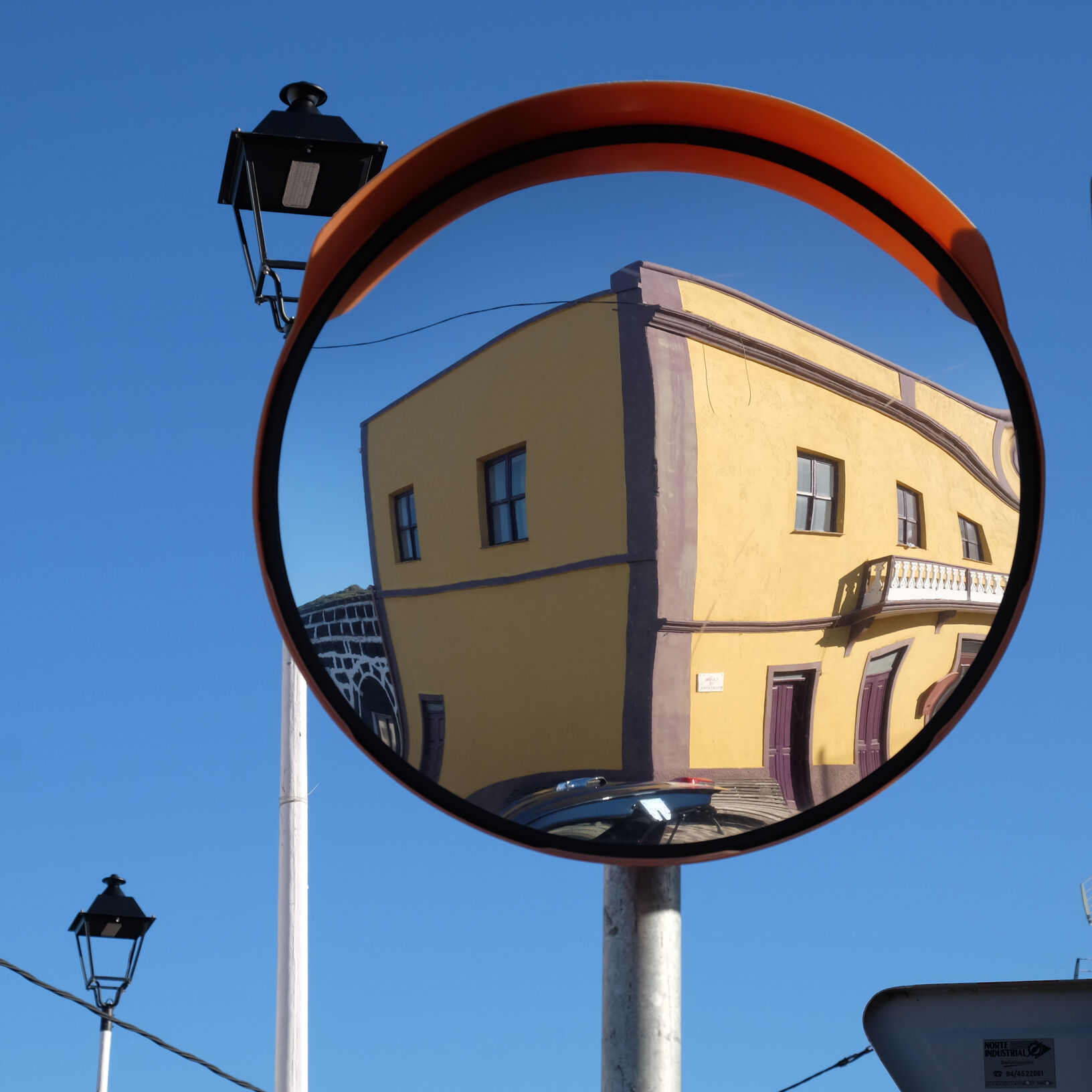} &
\includegraphics[width=0.3\textwidth]{resources/original/windmill.jpg}\\[0.5mm]
\small{\textit{garafia}} &
\small{\textit{mirror}} &
\small{\textit{windmill}}
\end{tabular}
\caption{Our six test images of size $3264\times 3264$. 
Photos by J.\ Weickert.}
\label{fig:images_dataset}
\end{figure}

\noindent {\bf Data Set and General Setting.}
We conduct experiments on a representative set of six natural images: 
\textit{boats}, \textit{elpaso}, \textit{flowers}, \textit{garafia}, 
\textit{mirror}, and \textit{windmill} (see \Cref{fig:images_dataset}). 
Our positive definite formulation \eqref{eq:formulation_projector} 
permits a conjugate gradient solver. Since it remains practical for 
positive definite systems even at large problem sizes, we can handle 
also high image resolutions. Therefore, all images have a resolution 
of $3264 \times 3264$. This is far higher than in most inpainting-based 
compression papers, with a  few exceptions such as~\cite{CW21,KCW25,SPKW23}.

%...........................................................................

%............................................................................

For each image, we perform $30$ densification iterations.
The sampling densities are chosen such that the reconstruction error 
remains perceptible, since we aim at a meaningful visual comparison 
between results for different feature types.

%----------------------------------------------------------------------------

\medskip
\noindent {\bf Comparability of Results for Different Feature Types.}
For classical inpainting with pointwise colour data in RGB
representation, the number of stored colour values is three times the 
number $m$ of mask pixels. 
In contrast, a corresponding Delaunay triangulation may contain up to $2m - 6$ 
triangles~\cite{AKL13}. Each triangle is associated with three average 
colour channel values. 
To ensure comparable data budgets across methods, we therefore employ lower 
mask densities for the Delaunay-based features. For classical inpainting 
we need to store an $x$- and $y$-coordinate for each mask point, 
and one value per colour channel. Under the simplifying assumption that 
each of these numbers requires equally many bits, the budget is proportional
to $5m$. For our Delaunay representation with $\widetilde{m}$ points, the 
data budget has an upper bound of $2\widetilde{m}+3(2\widetilde{m}-6)$. 
Balancing the budget of both feature types, we obtain the following:
For $m$ points in the pointwise colour data representation, we use 
$\widetilde{m} := \lfloor\frac{5m + 18}{8}\rfloor$ points in the 
Delaunay setting, where $\lfloor.\rfloor$ denotes the floor function.
Thus, $\,\widetilde{m} \approx \tfrac{5}{8}\,m\,$ for large $m$.

Obviously this formula can only serve as first guideline towards a fair 
comparison.
% As already mentioned, our manuscript does not discuss typical lossless
% or lossy storage strategies of the selected data. Many options are 
% possible there that can modify the formula in both directions. A
% detailed evaluation is beyond our scope. Instead we refer to the 
% journal paper of Mohideen et al.~\cite{MPW21}.
A rate-distortion evaluation under entropy coding is beyond the scope 
of the present paper and constitutes future work.

%-----------------------------------------------------------------------------

\begin{figure}[!tb]
\setlength{\tabcolsep}{2mm}
\centering
\begin{tabular}{ccc}
\small{\emph{flowers}, \,$3264 \times 3264$} & 
\small{pointwise, \,MSE: $90.10$} &
\small{Delaunay, \,MSE: $30.47$} \\[1mm]
\includegraphics[width=0.3\textwidth]
  {resources/original/flowers.jpg} &
\includegraphics[width=0.3\textwidth]
  {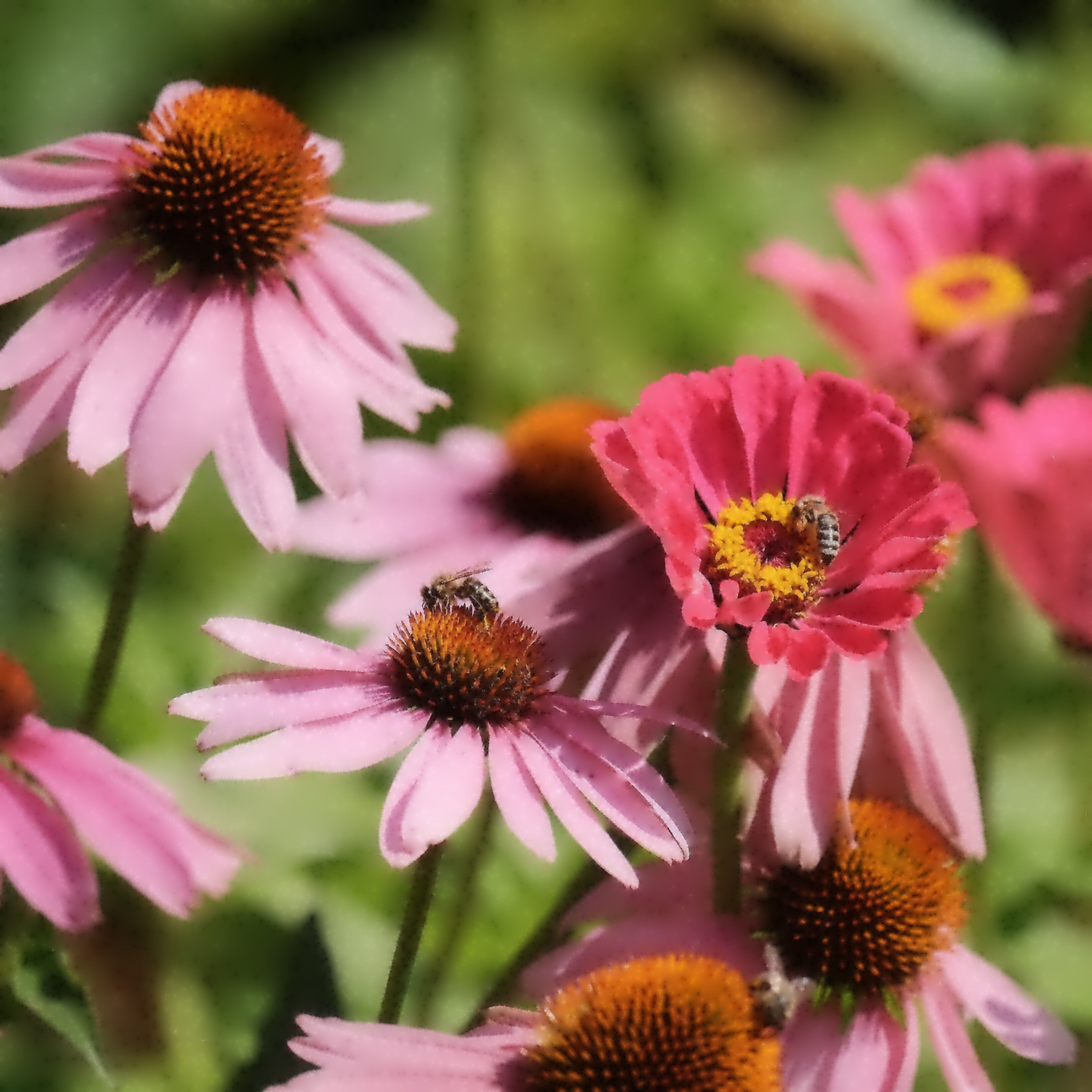} &
\includegraphics[width=0.3\textwidth]
  {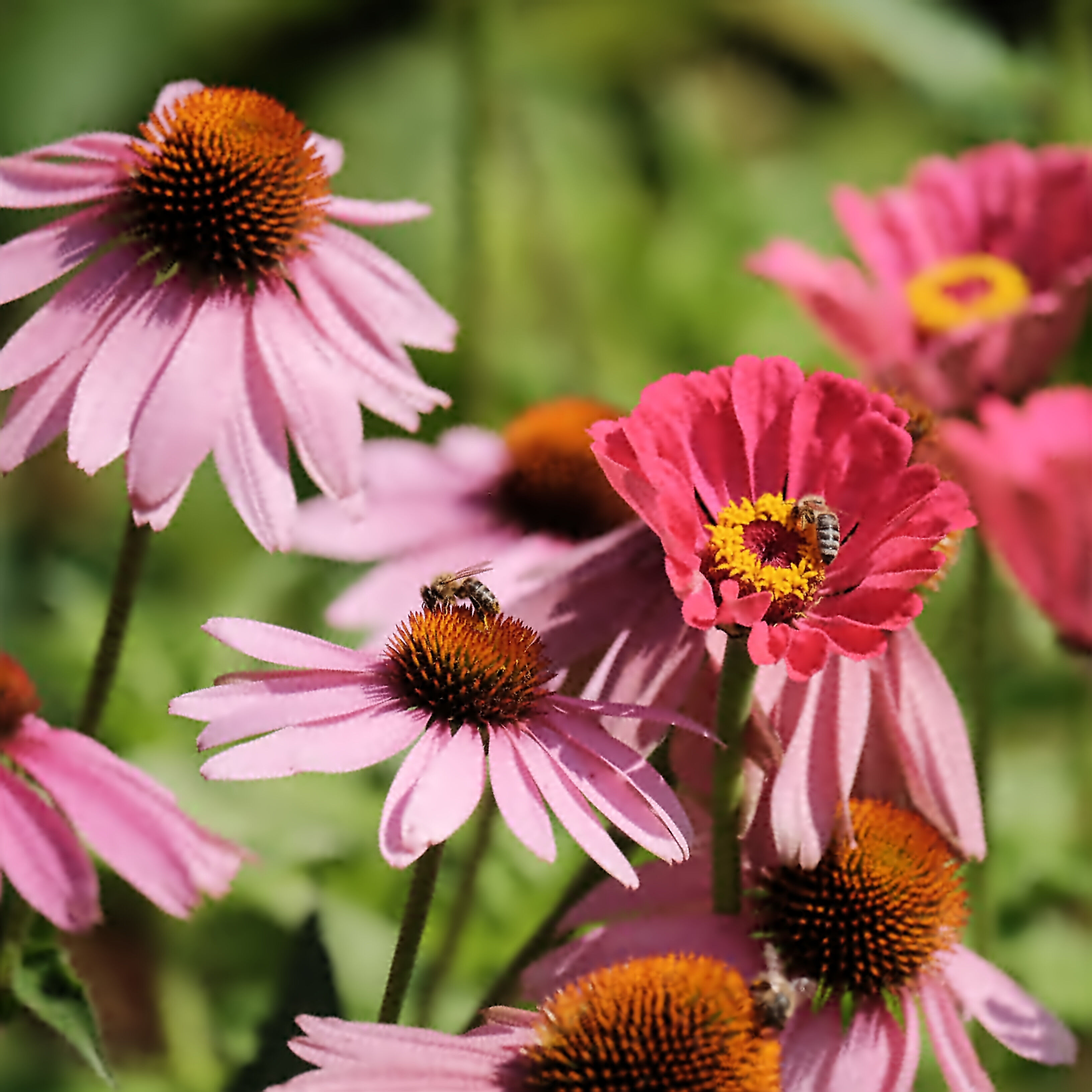}\\[3mm]
\includegraphics[width=0.3\textwidth]
  {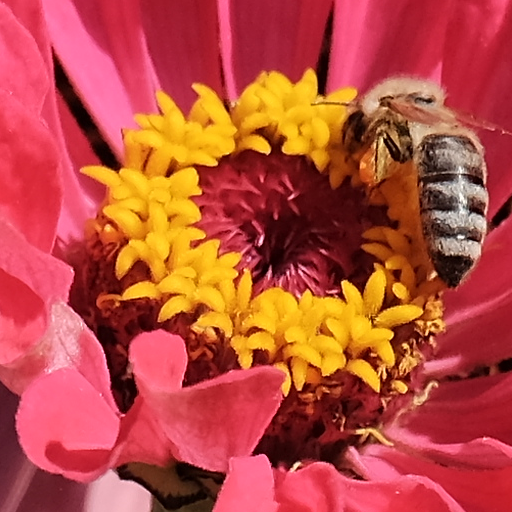} &
\includegraphics[width=0.3\textwidth]
  {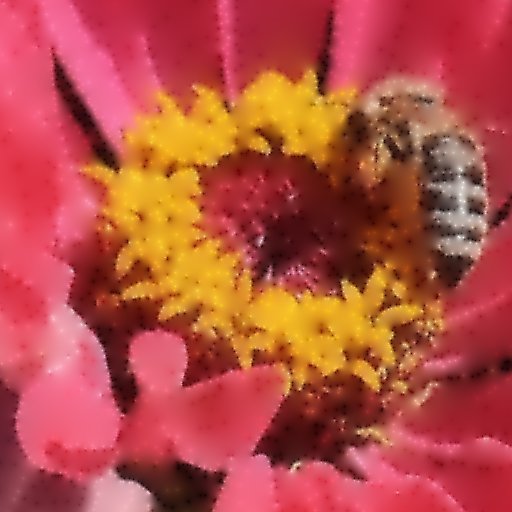} &
\includegraphics[width=0.3\textwidth]
  {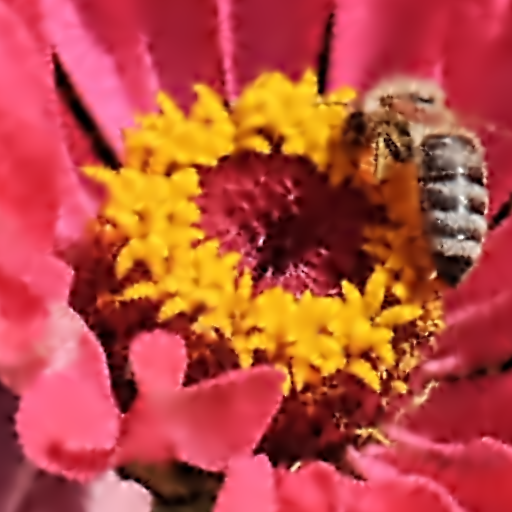}
\\[0.5mm]  
\small{\emph{flowers}, \,zoom} & 
\small{pointwise, \,zoom} &
\small{Delaunay, \,zoom} \\[1mm]
\end{tabular}
\caption{Comparison of the inpainting results for {\bf pointwise 
colour features at $\bm{0.3}$\,\% density versus our Delaunay-based features 
at $\bm{0.188}$\,\% density}. 
% The second row shows zoomed-in views of the same region for clarity. 
The pointwise reconstruction suffers from logarithmic singularities and 
a washed-out appearance. The Delaunay approach exhibits markedly improved 
contrast and edge connectivity. 
}
\label{fig:visual_comparison}
\end{figure}

\medskip
\noindent {\bf Quality Evaluation.}
\Cref{tbl:MSE_table_harmonic} reports the mean squared errors (MSEs) 
for the six images obtained with pointwise colour values and our 
Delaunay averages. Each RGB channel of the original images 
takes values in the range $[0, 255]$. We observe that inpainting with 
Delaunay averages consistently outperforms classical inpainting 
with pointwise values. On average, the MSE is reduced by 45.2 \%, 
leading to a remarkable PSNR improvement of 2.76 dB. 
For the image \textit{flowers}, the MSE almost drops by a factor three. 
\Cref{fig:visual_comparison} illustrates this improvement. 

%............................................................................

%............................................................................

\begin{table}
\caption{Mean squared errors (MSEs) for homogeneous diffusion inpainting 
using the pointwise colour values versus our proposed Delaunay averages. 
% The percentages refer to the mask densities for the pointwise colour values 
% setting. 
For the Delaunay case, lower mask densities are used to ensure a 
comparable data budget. See text for details.}

\label{tbl:MSE_table_harmonic}
\setlength{\tabcolsep}{2mm}
\begin{center}
\begin{tabular}{|l|c|c|c|c|c|c|c|c|}
\hline
Image & \textit{boats} & \textit{elpaso}  & \textit{flowers} 
& \textit{garafia} & \textit{mirror} & \textit{windmill} \\
\hline
Density Pointwise & $0.7\,\%$ & $0.3\,\%$ & $0.3\,\%$ & $0.7\,\%$ & $0.3\,\%$ 
        & $0.5\,\%$\\
Density Delaunay & $0.438\,\%$ & $0.188\,\%$ & $0.188\,\%$ & $0.438\,\%$ 
        & $0.188\,\%$ & $0.313\,\%$\\
\hline
MSE Pointwise & 184.98 & 81.31 & 90.10 &  176.38 & 56.29 & 191.97 \\
MSE Delaunay &  \textbf{119.76} & \textbf{50.43} & \textbf{30.47} 
& \textbf{116.93} & \textbf{22.18} & \textbf{120.07} \\
\hline 
\end{tabular}
\end{center}
\end{table} 

%% table with upscaled zoom of elpaso 
% \label{tbl:MSE_table_harmonic}
% \setlength{\tabcolsep}{2mm}
% \begin{center}
% \begin{tabular}{|l|c|c|c|c|c|c|c|c|}
% \hline
% Image & \textit{boats} & \textit{elpaso}  & \textit{flowers}
% & \textit{garafia} & \textit{mirror} & \textit{windmill} \\
% \hline
% Density Pointwise & $0.7\,\%$ & $0.3\,\%$ & $0.3\,\%$ & $0.7\,\%$ & $0.3\,\%$
%         & $0.5\,\%$\\
% Density Delaunay & $0.438\,\%$ & $0.188\,\%$ & $0.188\,\%$ & $0.438\,\%$
%         & $0.188\,\%$ & $0.313\,\%$\\
% \hline
% MSE Pointwise & 184.98 & 62.26 & 90.10 &  176.38 & 56.29 & 191.97 \\
% MSE Delaunay &  \textbf{119.76} & \textbf{42.00} & \textbf{30.47}
% & \textbf{116.93} & \textbf{22.18} & \textbf{120.07} \\
% \hline
% \end{tabular}
% \end{center}
% \end{table}

%-----------------------------------------------------------------------------

\begin{figure}
\setlength{\tabcolsep}{1.5mm}
\centering
\begin{tabular}{cccc}
\small{$408\times 408$} & 
\small{$816\times 816$} &
\small{$1632\times 1632$} &
\small{$3264\times 3264$}\\[1mm]
\includegraphics[width=0.225\textwidth]
   {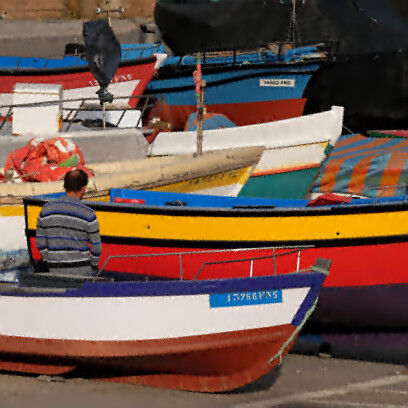} &
\includegraphics[width=0.225\textwidth]
   {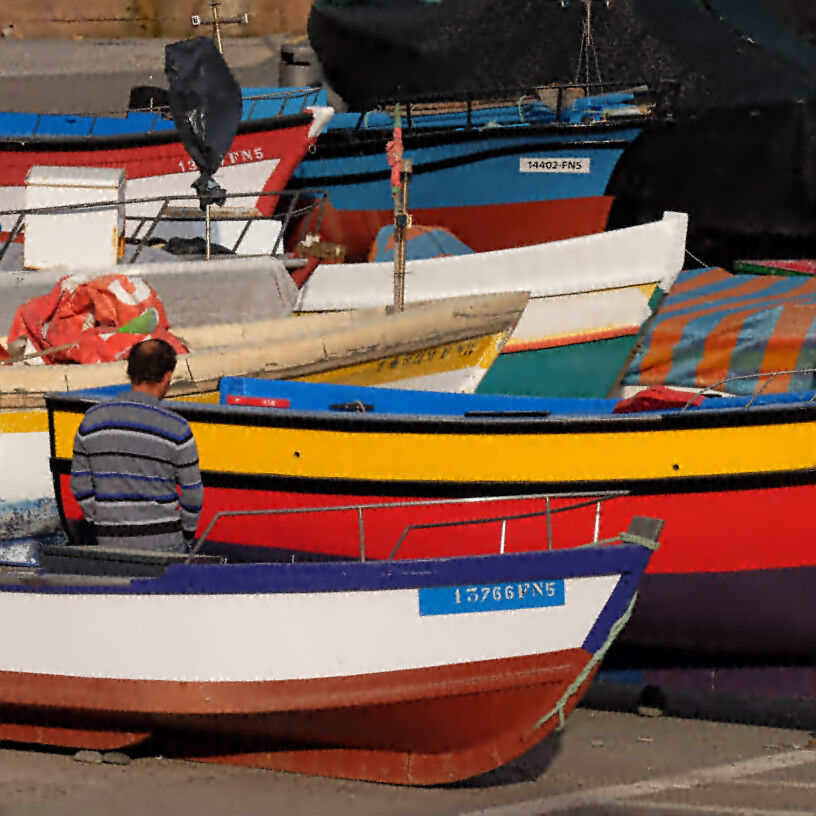} &
\includegraphics[width=0.225\textwidth]
   {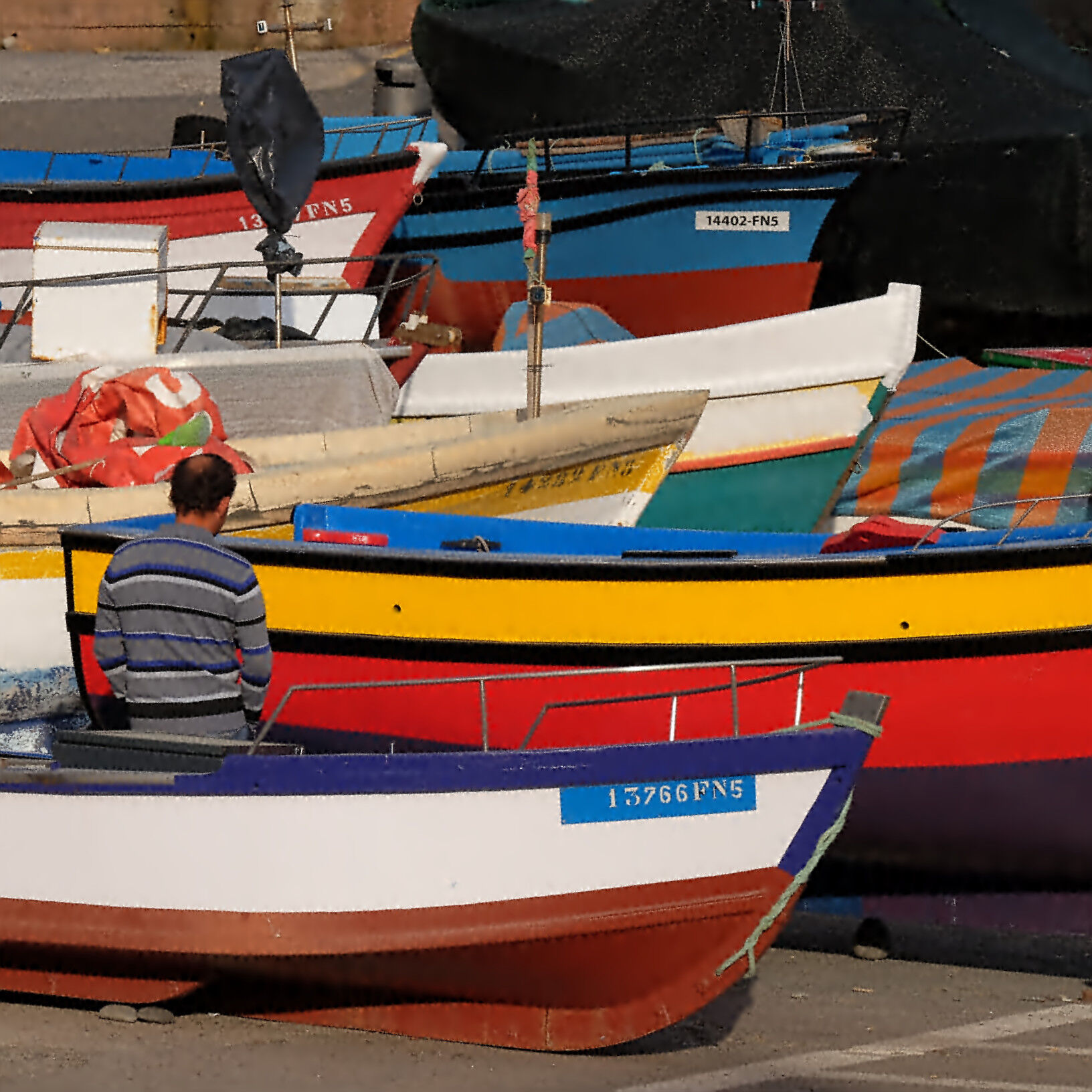} &
\includegraphics[width=0.225\textwidth] 
 {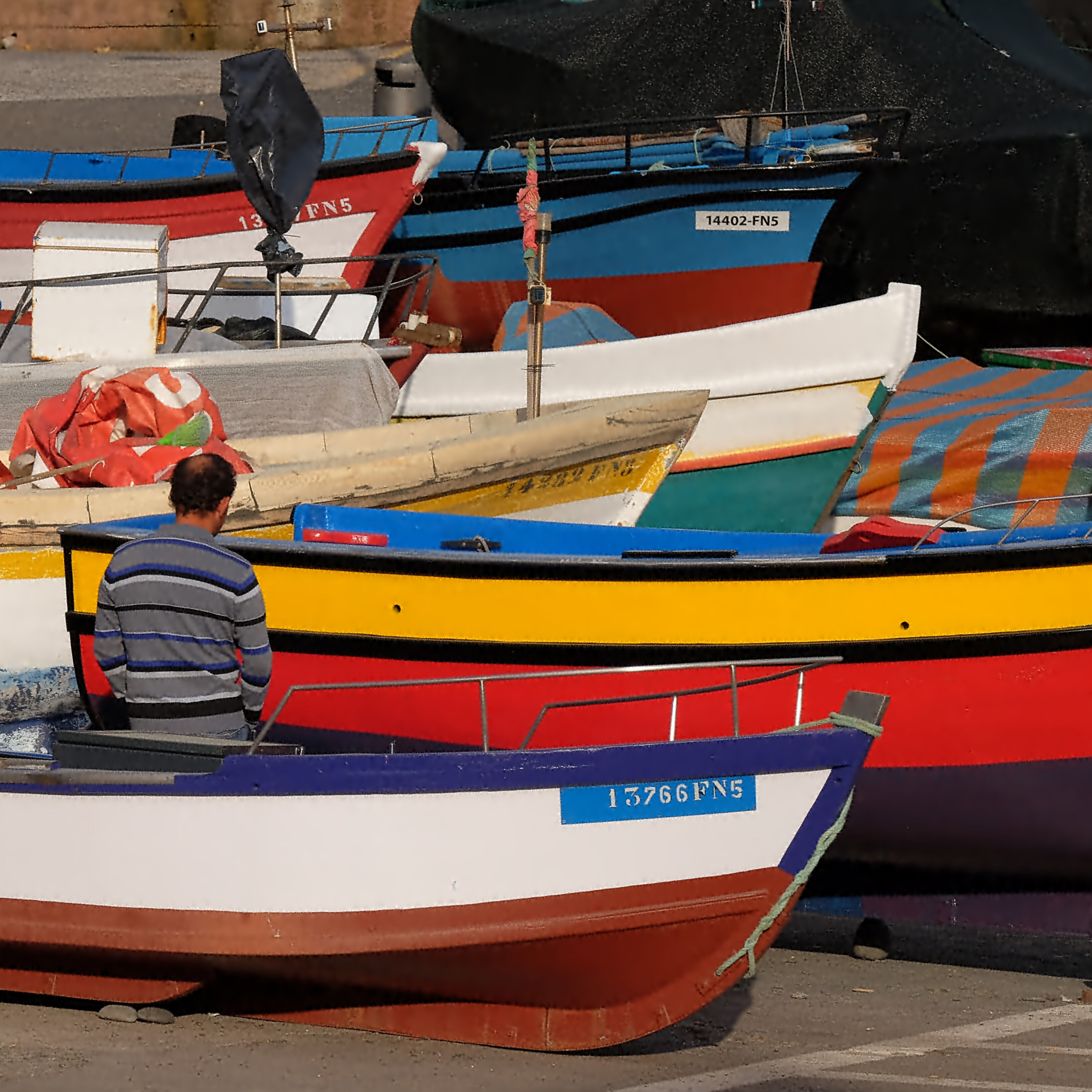}\\[1mm]
\small{$7.2\,\%$, MSE:$\,68.84$} &
\small{$3.6\,\%$, MSE:$\,72.26$} &
\small{$1.8\,\%$, MSE:$\,88.58$} &
\small{$0.9\,\%$, MSE:$\,73.33$} 
\end{tabular}
\caption{Doubling the resolution and halving the mask density maintains 
a similar MSE for the Delaunay features. The reported densities refer 
to the Delaunay vertices.}
\label{fig:scaling}
\end{figure}

\medskip
\noindent {\bf Scaling Behaviour.} 
Most publications on inpainting-based compression consider fairly
low resolution images and use mask densities around $5\,\%$.
Motivated by the surprisingly good reconstruction quality at 
densities below $1\%$, we investigate 
how appropriate mask densities scale with the image resolution. 
For our Delaunay feature, we find that doubling the 
image resolution in $x$- and $y$-direction allows to halve 
the mask density while maintaining a comparable MSE. \Cref{fig:scaling} 
illustrates this behaviour. A first intuition for this interesting 
observation may be found in Figure~\ref{fig:Delaunay}:
High feature densities arise typically near object or texture 
edges. These intrinsically 1-D structures grow linearly with the image 
resolution in $x$- and $y$-direction, whereas the pixel number is
area-based and hence grows quadratically. Thus, with increasing image 
resolution our approach allows higher compression rates -- without 
quality deteriorations. 

%-----------------------------------------------------------------------------

%%%%%%%%%%%%%%%%%%%%%%%%%%%%%%%%%%%%%%%%%%%%%%%%%%%%%%%%%%%%%%%%%%%%%%%%%%%%

\section{Conclusions and Outlook}
\label{sec:conclusions}

With our averages over Delaunay triangles, we have introduced the first 
adaptive feature type for inpainting-based image compression. It 
consistently outperformed common approaches based on pointwise colour 
values, with an average MSE improvement of 45.2 \%. It is remarkable 
that one can still achieve such significant progress after many years 
of research on data optimisation. This suggests that research on the 
feature type deserves more attention than the development of more 
sophisticated spatial and tonal data optimisation algorithms. While the 
latter was the focus of many papers, its improvements were more moderate.

The reason for the success of our single feature lies in its 
design: It adapts itself to the mask and combines efficient 
geometric representations (Delaunay triangulations) with robust analytical 
concepts (integral features). Moreover, we have developed a spatial 
optimisation strategy that is tailored specifically to this feature. 
It transfers ideas from stippling~\cite{DSZ17} to a new application field.

We have shown that homogeneous diffusion inpainting with Delaunay 
averages can be written in an elegant form that closely resembles the 
classical formulation. It creates a linear system with a positive definite 
system matrix. This guarantees the convergence of the classical 
conjugate gradient method. It is simple and can handle large images in 
practice. 

Interestingly, this goes hand in hand with another discovery: We have 
observed that the mask density can be halved when the resolution is 
doubled -- without compromising the approximation quality. This 
favourable scaling behaviour makes our approach particularly attractive 
for modern high-resolution imagery. Here it can produce appealing results 
even for data densities below {1\,\%}. 

We conjecture that our findings are of more general nature and 
usefulness:
Novel adaptive features -- single or multiple ones -- may give further
improvements, one can analyse different inpainting operators, other 
concepts from halftoning or stippling may be beneficial for data 
optimisation, and the favourable scaling behaviour may also hold for 
suitable alternative features and inpainting operators. Moreover, 
combining our representations with appropriate entropy coding 
strategies will lead to full codecs that allow comparisons to 
transform-based codecs. We are exploring these directions in our 
ongoing research.

%\subsubsection{Acknowledgements} Please place your acknowledgments at
%the end of the paper, preceded by an unnumbered run-in heading (i.e.
%3rd-level heading).

%%%%%%%%%%%%%%%%%%%%%%%%%%%%%%%%%%%%%%%%%%%%%%%%%%%%%%%%%%%%%%%%%%%%%%%%%%%

% ---- Bibliography ----
%
% BibTeX users should specify bibliography style 'splncs04'.
% References will then be sorted and formatted in the correct style.
%
\bibliographystyle{splncs04}
\bibliography{myrefs, additional_refs}
\end{document}